\documentclass[a4paper,11pt]{article}
\usepackage{pos}

\usepackage{adjustbox}
\usepackage{amsmath}
\usepackage{graphicx}
\usepackage{booktabs}
\usepackage{subfigure}

\title{Electron Scattering at the Intensity Frontier with SoLID}

\author*[a]{Zein-Eddine Meziani (for the SoLID Collaboration)}
\affiliation[a]{Argonne National Laboratory,\\
 9700 South Cass Avenue, Lemont, USA}

\emailAdd{zmeziani@anl.gov}

\abstract{The Solenoidal Large Intensity Device (SoLID) is a large acceptance spectrometer capable of operating at the luminosity frontier. It is proposed to fully exploit the scientific potential of the continuous electron beam accelerator facility (CEBAF) 12 GeV energy upgrade at Jefferson Lab. Its conceptual design is mature, having passed multiple reviews and been validated by a successful pre-R\&D phase. The envisioned scientific program consists of three avenues of research, namely the  3D momentum imaging of the structure of the nucleon, the origin of the proton mass through the gluonic gravitational form factors (GFFs), and the search of physics beyond the standard model of particle physics. These avenues are complemented by a growing supplemental list of run group experiments that address a variety of important topics.}

\FullConference{31st International Workshop on Deep Inelastic Scattering (DIS2024)\\
 8–12 April 2024\\
Grenoble, France\\}

\begin{document}
\maketitle

\section{Introduction; SoLID "raison d'\^etre"}

To address some of the key questions in hadronic and nuclear physics highlighted in the Long Range Plan of Nuclear Physics~\cite{Dodge:2024}, CEBAF's 12 GeV upgrade at Jefferson Lab has enabled significant electron scattering data collection over the last five years with four operational experimental halls~\cite{Arrington:2021alx}. On one hand, Halls B  and  D  are equipped with large-acceptance, toroidal (CLAS12) and solenoidal (GlueX) spectrometers respectively, operating at moderate luminosities about 10$^{35}$ cm$^{-2}$s$^{-1}$. On the other hand, Halls A and C are equipped with spectrometers (HRS, SBS, HMS, SHMS) that offer high to moderate momentum resolution and can reach a maximum central momentum of up to 11 GeV. Although these spectrometers have a smaller acceptance, they can operate at significantly higher luminosity up to 10$^{39}$ cm$^{-2}$s$^{-1}$. In contrast and to fully leverage the 12 GeV energy upgrade of CEBAF, the proposed Solenoidal Large Intensity Device (SoLID) is essential  to enable a large acceptance detector to handle a luminosity in the range 10$^{37}$-10$^{39}$ cm$^{-2}$s$^{-1}$. The SoLID experiment is unique in that it operates at the luminosity frontier to enable multidimensional measurements in hadronic/nuclear physics. SoLID's  features enable the most challenging measurements to be performed with high precision in the valence quark region of nucleons and nuclei. SoLID's scientific program and detector are described in detail in its white paper Ref.~\cite{JeffersonLabSoLID:2022iod}.

\section{SIDIS in SoLID and TMDs}

\begin{figure}[h!]
    \centering
    \includegraphics[width=0.7\linewidth]{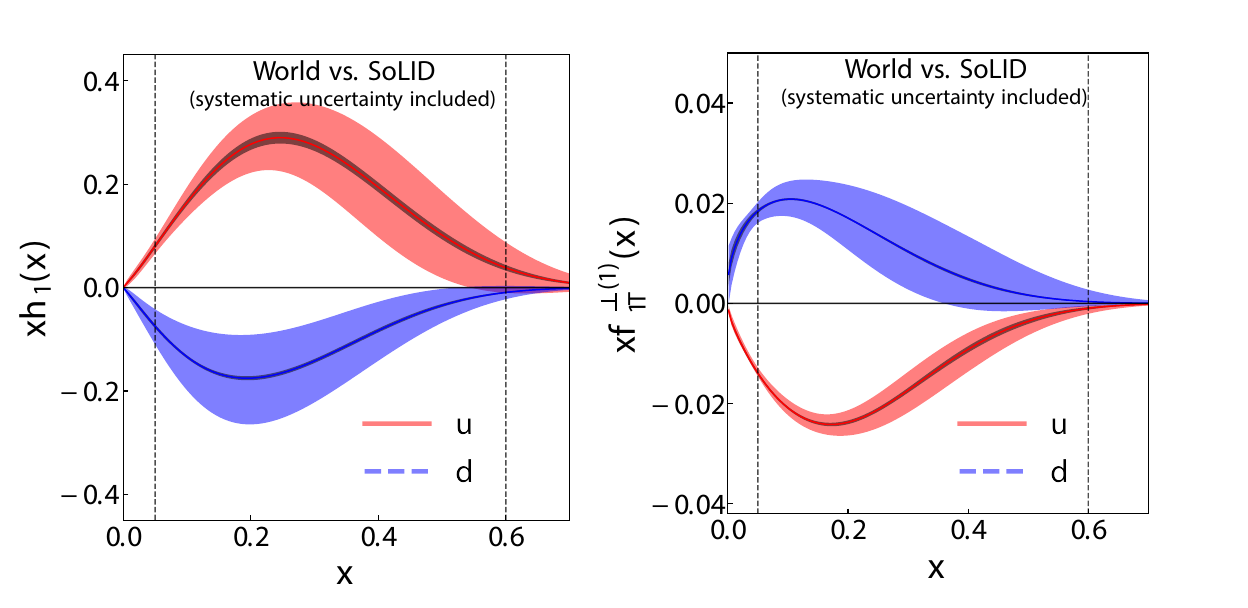}
    \caption{SoLID uncertainties projections on the transversity $xh_1(x)$ (left) and Sivers  $xf^{\perp(1)}_{1T}$(right) distributions as a function of $x$ for the up (red) and down (blue) quarks compared to the world data extraction (wide bands).}
    \label{fig:tmd}
\end{figure}
The transverse parton momentum structure of the nucleon is rich but its unraveling has taken center stage only in the last 20 years. One way to access the missing information, namely the nucleon partons' transverse momentum is through semi-inclusive deep inelastic scattering (SIDIS)~\cite{Boussarie:2023izj}.
SoLID will advance the 3D momentum imaging of the partonic structure of the nucleon by providing high precision data for each bin of multi-variables relevant to the SIDIS experiments~\cite{JeffersonLabSoLID:2022iod} in the valence quark region.  As an example, Fig.~\ref{fig:tmd} shows the impact of proposed SoLID SIDIS measurements on the extraction of the transversity and Sivers functions compared to the present world data.

\section {Near-Threshold $J/\psi$ Production in SoLID and gluonic GFFs}
\begin{figure}[h!]
    \centering
    \includegraphics[width=0.6\linewidth]{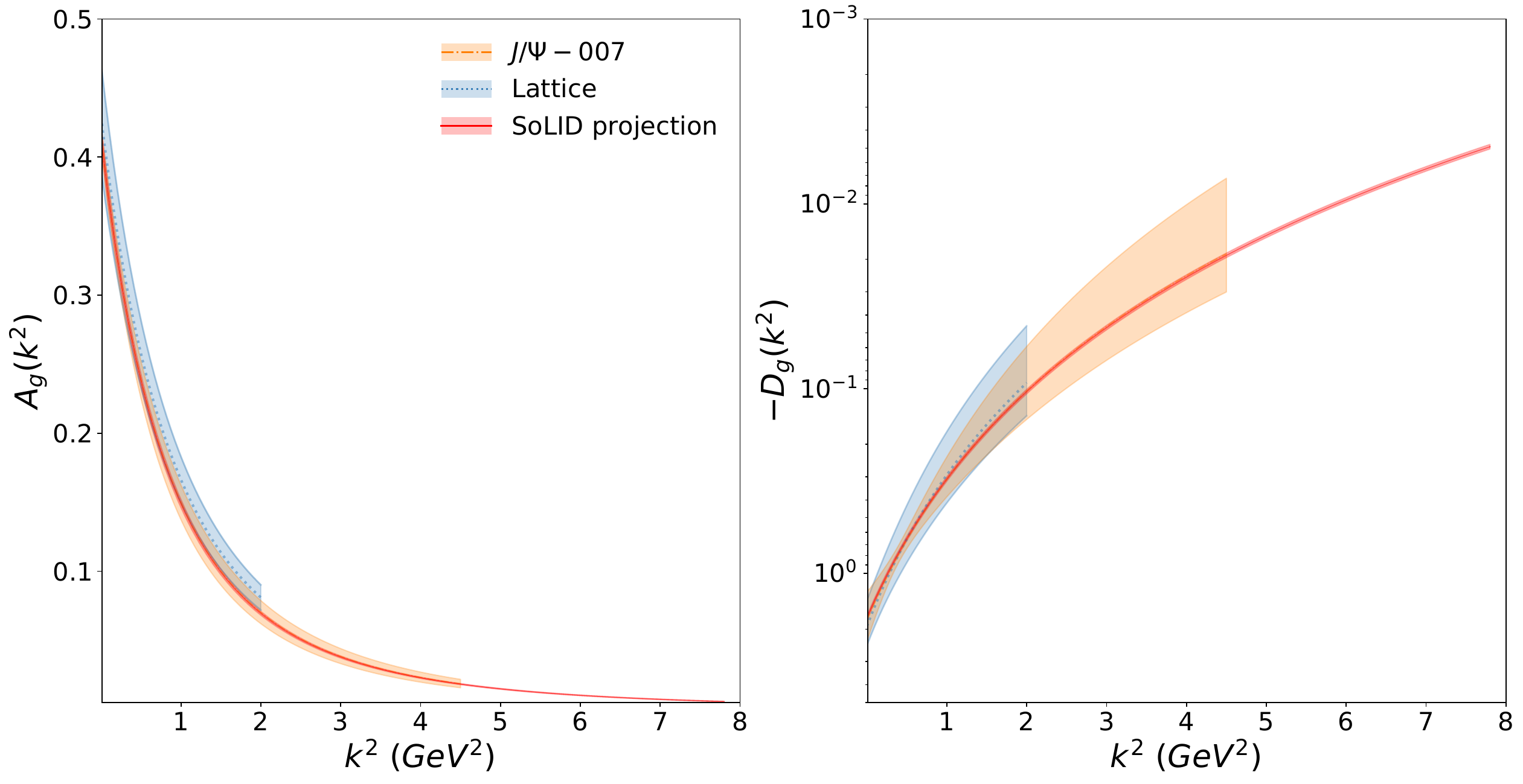}
   \includegraphics[width=0.36\linewidth]{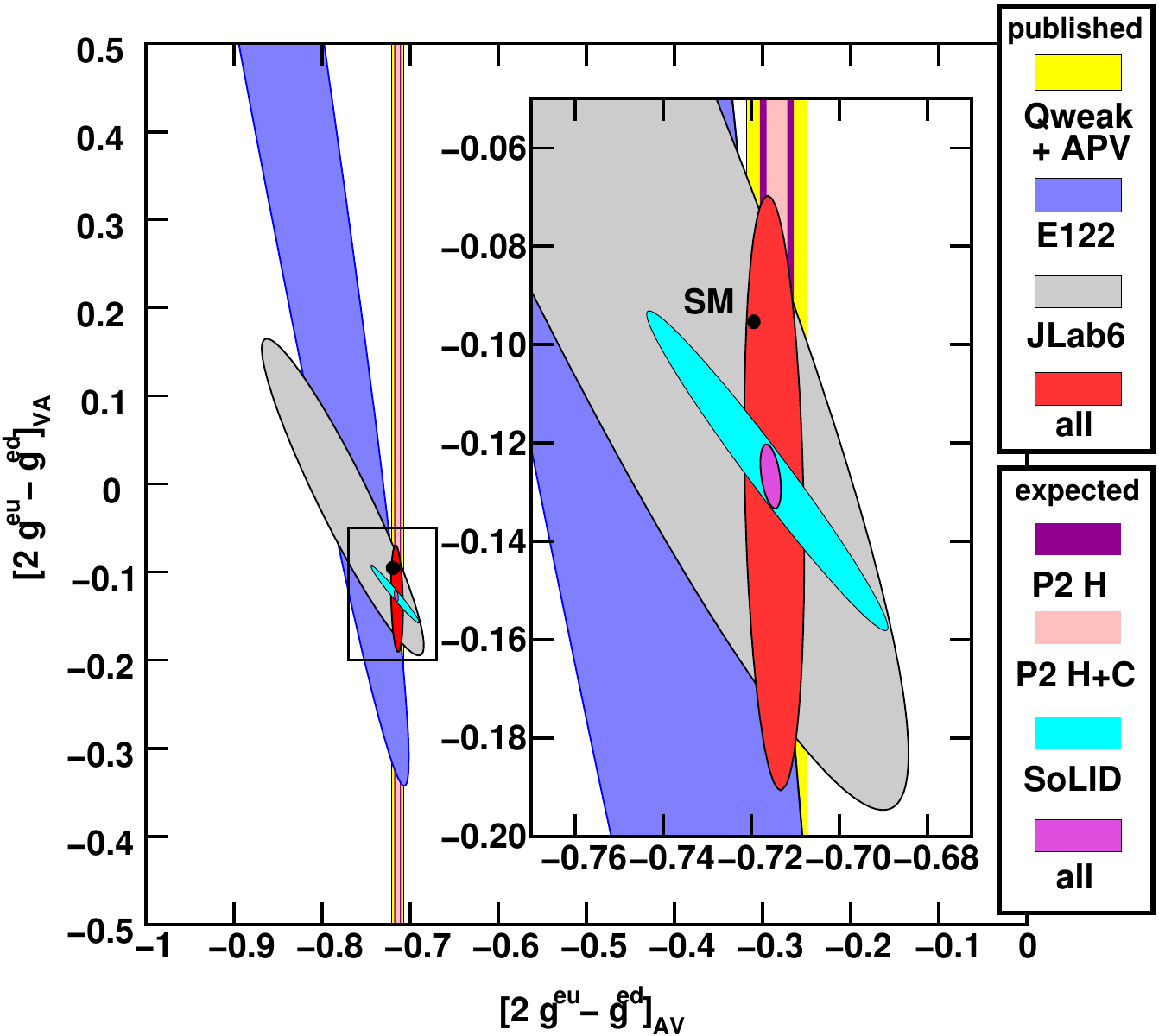}
    \caption{Left two panels, gluonics GFFs $A_g(k)$ and $D_g(k)$ extracted from the SoLID projected data and compared to $J/\psi-007$ current determination and lattice QCD. Right panel, impact of the SoLID PVDIS experiment on the determination of the couplings $g_{VA}^{eq}$ (vertical axis) versus $g_{AV}^{eq}$  (horizontal axis)~\cite{JeffersonLabSoLID:2022iod,Zheng:2021hcf}.}
    \label{fig:gff-pvdis}
\end{figure}

Recently, it was suggested that the measurement of near threshold photo-production of $J/\psi$ on the proton would give access to fundamental gluonic mechanical properties of the proton described by the energy-momentum form factors~\cite{Kobzarev:1962wt,Pagels:1966zza}. Measurements of such a reaction where performed by GlueX~\cite{GlueX:2023pev}, $J/\psi-007$~\cite{Duran:2022xag} and CLAS12~\cite{CLAS12}. In particular in $J/\psi-007$, the authors extracted, for the first time, the gluon contribution of the gravitational form factors (GFFs) $A_g(k)$ and $D_g(k)$. Studies on the quark sector are related to the measurements of the Compton form factors (CFFs) through deeply virtual Compton Scattering (DVCS) and the extraction of the generalized parton distributions(GPD). In this case and the for the first time, $D_q(k)$ quark gravitational form factor was extracted~\cite{Burkert:2018bqq}. Both $A(k)$ and $D(k)$ are needed to discuss the mass density distribution in the proton. However, with knowledge of $D(k)$ alone, the pressure density distribution in the proton has been determined~\cite{Burkert:2018bqq}. In Fig.~\ref{fig:gff-pvdis} we show the projected impact of the SoLID $J/\psi$ experiment compared to the current determination of the gluonic form factors including those of lattice quantum chromodynamics (QCD)~\cite{Pefkou:2021fni}.

\section{SoLID Parity Violation Deep Inelastic Scattering Experiment}
Parity violation in deep inelastic scattering (PVDIS) in search of physics beyond the standard model is the most demanding measurement in SoLID and requires its own configuration of detectors to accept an unprecedented luminosity in a large acceptance detector, about 10$^{39}$ cm$^{-2}$s$^{-1}$. More details on the challenges and impact of this experiment on physics beyond the standard model search as well hadronic structure of the proton, like the $x$ dependence of the ratio of down- over up-quark distribution functions are discussed in Ref.~\cite{JeffersonLabSoLID:2022iod}. SoLID with its deuteron PVDIS measurement will provide new constraints on the effective couplings $g_{VA}^{eq}$ and $g_{AV}^{eq}$, and unique sensitivity to BSM physics, see Fig.~\ref{fig:gff-pvdis} (most right panel)  adapted from \cite{Zheng:2021hcf}.
Together with the upcoming MOLLER~\cite{MOLLER:2014iki} measurement at JLab and the P2 experiment~\cite{Becker:2018ggl} at the MESA facility in Mainz,  SoLID PVDIS will also map out the weak mixing angle $\sin^2(\theta_W)$ in the low to intermediate energy region, complementary to BSM search provided by the electron-quark couplings.

\section{Summary}
SoLID offers an exceptional and unique opportunity to enhance and extend the scientific reach of the 12 GeV upgrade of CEBAF at Jefferson Lab. For more complete description of the SoLID science program and detector configurations details we refer the reader to Ref.~\cite{JeffersonLabSoLID:2022iod}


\begin{thebibliography}{99}

\bibitem{Dodge:2024}
Dodge, G. E., 
\href{https://doi.org/10.1080/10619127.2024.2303306}
{\emph{(2024). The U.S. Nuclear Science Long Range Plan. Nuclear Physics News,} 34(1), 3–4}

\bibitem{JeffersonLabSoLID:2022iod}
J.~Arrington, \textit{et al.} [Jefferson Lab SoLID],
\emph{The solenoidal large intensity device (SoLID) for JLab 12 GeV},
\href{https://doi.org/10.1088/1361-6471/acda21}
{\emph{J. Phys. G} \textbf{50} (2023) no.11, 110501}
[{\tt arXiv:2209.13357 [nucl-ex]}].

\bibitem{Arrington:2021alx}
J.~Arrington,\textit{et al.},
{\emph{Physics with CEBAF at 12 GeV and future opportunities}},
\href{https://doi.org/10.1016/j.ppnp.2022.103985}
{\emph{Prog. Part. Nucl. Phys.} \textbf{127} (2022), 103985}
[{\tt arXiv:2112.00060 [nucl-ex]}].

\bibitem{Boussarie:2023izj}
R.~Boussarie, \textit{et al.},
\emph{TMD Handbook},
[{\tt arXiv:2304.03302 [hep-ph]}].

\bibitem{Kobzarev:1962wt}
I.~Y.~Kobzarev and L.~B.~Okun,
\emph{Gravitational interactions of Fermions,} 
{\emph{Zh. Eksp. Teor. Fiz.} \textbf{43}, 1904-1909 (1962)}

\bibitem{Pagels:1966zza}
Pagels, H., \emph{Energy-Momentum Structure Form Factors of Particles,} 
\href{https://doi:10.1103/PhysRev.144.1250.} 
{\emph{Phys. Rev.} \textbf{144}, (1966)1250--1260}

\bibitem{GlueX:2023pev}
S.~Adhikari \textit{et al.} [GlueX],
\emph{Measurement of the J/$\psi$ photoproduction cross section over the full near-threshold kinematic region,} 
\href{https://doi.org/10.1103/PhysRevC.108.025201} 
{\emph{Phys. Rev. C} \textbf{108}, no.2, 025201 (2023) }
[{\tt arXiv:2304.03845 [nucl-ex]}].

\bibitem{Duran:2022xag}
B.~Duran, \textit{et al.}
\emph{Determining the gluonic gravitational form factors of the proton,}
\href{https://doi.org/10.1038/s41586-023-05730-4}
{\emph{Nature} \textbf{615} (2023) no.7954, 813}
[{\tt arXiv:2207.05212 [nucl-ex]}].

\bibitem{CLAS12}
P. Nadel-Turonski \& S. Stepanyan  (contacts) \textit{et al.}, 
\emph{Timelike Compton Scattering and J/psi photoproduction on the proton in $e^+e^-$ pair production with CLAS12 at 11 GeV,} 
\href{https://misportal.jlab.org/mis/physics/experiments/searchProposals.cfm?paramHall=B&paramExperimentEnergy=12GeV&paramExperimentStatusList=A,G,H}
{\emph{E12-12-001A}}. 

\bibitem{Burkert:2018bqq}
V.~D.~Burkert, L.~Elouadrhiri and F.~X.~Girod,
\emph{The pressure distribution inside the proton},
\href{http://doi.org/10.1038/s41586-018-0060-z}
{\emph{Nature} \textbf{557} (2018) no.7705, 396-399}

\bibitem{Pefkou:2021fni}
D.~A.~Pefkou, D.~C.~Hackett and P.~E.~Shanahan,
\emph{Gluon gravitational structure of hadrons of different spin},
\href{https://doi.org/10.1103/PhysRevD.105.054509}
{\emph{Phys. Rev. D} \textbf{105} (2022) no.5, 054509}
[{\tt arXiv:2107.10368 [hep-lat]}].

\bibitem{MOLLER:2014iki}
J.~Benesch \textit{et al.} [MOLLER],
\emph{The MOLLER Experiment: An Ultra-Precise Measurement of the Weak Mixing Angle Using M{\"o}ller Scattering}, [{\tt arXiv:1411.4088 [nucl-ex]}].

\bibitem{Becker:2018ggl}
D.~Becker, \textit{et al.}
\emph{The P2 experiment},
\href{https://doi.org/10.1140/epja/i2018-12611-6}
{\emph{Eur. Phys. J. A} \textbf{54} (2018) no.11, 208}
[{\tt arXiv:1802.04759 [nucl-ex]}].

\bibitem{Zheng:2021hcf}
X.~Zheng, J.~Erler, Q.~Liu and H.~Spiesberger,
\emph{Accessing weak neutral-current coupling $g_{AA}^{eq}$ using positron and electron beams at Jefferson Lab},
\href{https://doi.org/10.1140/epja/s10050-021-00490-z}
{\emph{Eur. Phys. J. A} \textbf{57} (2021) no.5, 173}
[{\tt arXiv:2103.12555 [nucl-ex]}].

\end{thebibliography}
\end{document}